\documentclass[sigconf, authorversion]{acmart}

\usepackage[table]{xcolor}
\usepackage{hyperref}
\usepackage{url}
\usepackage{tcolorbox}
\usepackage{graphicx}
\usepackage{float}
\usepackage{threeparttable}
\usepackage{siunitx}
\usepackage{soul} 
\usepackage{tabularx}
\usepackage{ragged2e}
\usepackage{stfloats}

\AtBeginDocument{%
  }

\setcopyright{acmlicensed}
\copyrightyear{2025}
\acmYear{2025}
\acmDOI{10.48550/arXiv.2504.14345} 
\acmConference[CIKM'25 Workshop on FinAI]{Advances in Financial AI}{November 14, 2025}{Seoul, Republic of Korea}
\acmBooktitle{Proceedings of the 34th ACM International Conference on Information and Knowledge Management (CIKM '25), November 10--14, 2025, Seoul, Republic of Korea}
\acmISBN{978-1-4503-XXXX-X/2025/11} 

\begin{document}

\title{LLM-Enhanced Black-Litterman Portfolio Optimization}

\author{Youngbin Lee}
\authornote{These authors contributed equally to this work.}
\email{youngandbin@elicer.com}
\affiliation{%
  \institution{Elice \\ AI Quant Lab, MODULABS}
  \city{Seoul}
  \country{Republic of Korea}
}

\author{Yejin Kim}
\authornotemark[1]
\email{yejin.kim.ds@meritz.com}
\affiliation{%
  \institution{Meritz Fire \& Marine Insurance \\ AI Quant Lab, MODULABS}
  \city{Seoul}
  \country{Republic of Korea}
  }

\author{Juhyeong Kim}
\email{juhyeong.kim@miraeasset.com}
\email{nonconvexopt@gmail.com}
\affiliation{%
  \institution{Mirae Asset Global Investments \\ AI Quant Lab, MODULABS}
  \city{Seoul}
  \country{Republic of Korea}
  }

\author{Suin Kim}
\email{suin@elicer.com}
\affiliation{%
  \institution{Elice}
  \city{Seoul}
  \country{Republic of Korea}}

\author{Yongjae Lee}
\authornote{Corresponding author}
\email{yongjaelee@unist.ac.kr}
\affiliation{%
  \institution{Ulsan National Institute of Science and Technology}
  \city{Ulsan}
  \country{Republic of Korea}}

\renewcommand{\shortauthors}{Lee et al.}

\definecolor{hotpink}{rgb}{1,0.41,0.71}
\begin{abstract}
The Black-Litterman model addresses the sensitivity issues of traditional mean-variance optimization by incorporating investor views, but systematically generating these views remains a key challenge.
This study proposes and validates a systematic framework that translates return forecasts and predictive uncertainty from Large Language Models (LLMs) into the core inputs for the Black-Litterman model: investor views and their confidence levels.
Through a backtest on S\&P 500 constituents, we demonstrate that portfolios driven by top-performing LLMs significantly outperform traditional baselines in both absolute and risk-adjusted terms. 
Crucially, our analysis reveals that each LLM exhibits a distinct and consistent investment style
which is the primary driver of performance.
We found that the selection of an LLM is therefore not a search for a single best forecaster, but a strategic choice of an investment style whose success is contingent on its alignment with the prevailing market regime.
The source code and data are available at \textcolor{hotpink}{\url{https://github.com/youngandbin/LLM-BLM}}.
\end{abstract}





\maketitle

\section{Introduction}

Portfolio optimization is a fundamental problem in finance, traditionally approached using mean-variance optimization \cite{markowitz1952jf}.
However, a well-known limitation of mean-variance optimization is its sensitivity to input parameters, such as expected returns and covariance estimates. 
The Black-Litterman asset allocation model \cite{black1992global} was introduced to mitigate these problems by incorporating investor views into expected returns through a Bayesian framework. 
Unlike traditional mean-variance optimization, which relies purely on historical data, the Black-Litterman model blends prior market equilibrium returns with subjective investor views, weighted by confidence levels. 
This approach not only stabilizes return estimates but also allows for a more flexible and systematic integration of expert knowledge and quantitative signals. While theoretical research has extended the framework to complex dynamic, multi-period settings \cite{abdelhakmi2025dynamic}, our study utilizes the conventional single-period model to focus specifically on the novel challenge of generating views from LLMs.

Although the Black-Litterman model remains widely used in asset management industry, including Goldman Sachs, a key challenge lies in how to define and quantify these views in a systematic and reliable manner. 
Traditionally, these views have been derived from qualitative assessments or ad hoc quantitative models, limiting scalability and objectivity. 

This study proposes a systematic framework for integrating LLM-generated views into the Black-Litterman model. Our contribution lies not in developing a new time-series forecasting model with LLMs, but in \textbf{establishing a structured methodology to translate LLM predictions into actionable portfolio views.} 
This framework systematically incorporates not only the LLM's return predictions but also their inherent uncertainty, allowing for a data-driven and scalable generation of views while mitigating human bias. 
Our experiments serve to illustrate how this framework operates in practice, demonstrating that the final portfolio performance becomes contingent on the quality of the views generated by the specific LLM employed.

The primary objective of this research is to empirically demonstrate our proposed framework. We apply it to generate portfolios based on views from several state-of-the-art LLMs, using S\&P 500 constituents as the asset universe. 
We use a dedicated validation period to tune the Black-Litterman model's key hyperparameter, $\tau$, which governs the confidence in market equilibrium returns versus LLM-generated views. The optimized portfolio is then backtested on an unseen test period against traditional benchmarks. 
The analysis focuses not on crowning a superior forecasting model, but on illustrating how the characteristics of different LLM-generated views, when systematically integrated via our framework, lead to distinct portfolio performance outcomes. Performance is assessed using standard metrics such as the Sharpe ratio, providing a clear picture of how each model's predictive patterns translate into investment results.

\section{Related Works}

\subsection{The Black-Litterman Model and the Challenge of View Generation}

The Black-Litterman model \cite{black1992global} offered a novel approach to portfolio optimization by systematically blending market equilibrium returns with investor-specific views. 
A key advantage of this model is that it produces more intuitive and stable portfolios than traditional mean-variance optimization \cite{he2002intuition, meucci2010black}. 

However, the practical application of the model has always been limited by its most critical input: the generation of investor views.
Historically, these views were based on the subjective judgments of human experts \cite{idzorek2007step}. This process is difficult to scale and is prone to cognitive biases, such as overconfidence and herding, which are widely studied in behavioral finance \cite{barberis2003survey, kara2019hybrid}. The challenge of converting expert opinions into a quantitative format while avoiding human error has led to a long search for a more systematic and data-driven approach to view generation.


To address the subjectivity of human views, researchers turned to machine learning, which has shown significant success in empirical asset pricing \cite{gu2020empirical}. Early approaches employed quantitative models to forecast returns based on historical patterns, using techniques like temporal relational ranking or multi-graph convolutional networks \cite{feng2019temporal, wang2022mg}.

More sophisticated methods have been developed to systematically generate views. For instance, machine learning has been used to systematically generate views based on established asset pricing factors \cite{ko2025portfolio}. Other studies have integrated advanced deep learning architectures like Transformers and Generative Adversarial Networks (GANs) to capture complex, non-linear market dynamics and refine predictions \cite{zhu2024enhancing}. Beyond prediction, some research has explored deep reinforcement learning to directly learn optimal trading policies \cite{jiang2017deep}. While these methods are a significant step towards objectivity, they share a common limitation: they are primarily designed for structured inputs. This means that processing unstructured information, such as natural language, requires extensive feature engineering. Consequently, their ability to adapt to new information sources or changing market regimes remains limited.

\subsection{LLMs in Finance}

Recent advances in LLMs have created new opportunities in finance, particularly due to their ability to understand and reason with unstructured text data. This has led to the development of specialized financial models like BloombergGPT \cite{wu2023bloomberggpt} and open-source initiatives like FinGPT \cite{liu2023fingpt}. As a result, LLMs are being applied to a wide range of fianncial tasks, as surveyed by \cite{lee2024survey}.

The application of LLMs in finance now extends far beyond simple text summarization. 
For example, studies have used LLMs to generate qualitative evaluations for asset pricing models \cite{cheng2024empirical}, analyze news sentiment for return forecasting \cite{lopez2024can}, and create specialized financial asset representations to overcome the limitations of general-purpose embeddings \cite{lee2025theme}. These diverse capabilities highlight the potential of LLMs to serve as powerful signal generators for portfolio optimization \cite{zhao2024revolutionizing, hwang2025decision}.

Another key application area is return forecasting. Studies have explored using LLMs for return forecasting \cite{jin2023time}, with techniques like retrieval-augmented generation (RAG) further improving prediction accuracy \cite{xiao2025retrieval}. However, a critical gap remains: these studies often focus on the accuracy of the forecast itself, rather than on the crucial subsequent step of how to systematically integrate these powerful, yet inherently uncertain, predictions into a robust portfolio construction framework. Simply using raw LLM outputs can lead to unstable portfolios, as LLMs possess their own inherent biases that can distort recommendations if left unchecked \cite{lee2025your}. A separate line of research even bypasses optimization frameworks entirely, designing fully autonomous trading agents where the LLM is the sole decision-maker \cite{xiao2024tradingagents, yu2025finmem}.

Our work bridges the gap between unstructured LLM insights and structured portfolio optimization. We move beyond merely generating forecasts and propose a systematic framework that translates not only the return predictions from LLMs but also their inherent uncertainty into the core inputs of the Black-Litterman model. By quantifying this uncertainty, our approach automatically calibrates the confidence in LLM-generated views, mitigating the risk of overweighting biased or unstable predictions. To our knowledge, this is the first study to establish a structured methodology for integrating both the predictive power and the predictive uncertainty of various LLMs into a proven portfolio optimization model, thereby addressing the challenge of systematic view generation in a novel and scalable manner.

\section{Method}

\begin{figure*} [h!]
    \includegraphics[width=1\linewidth]{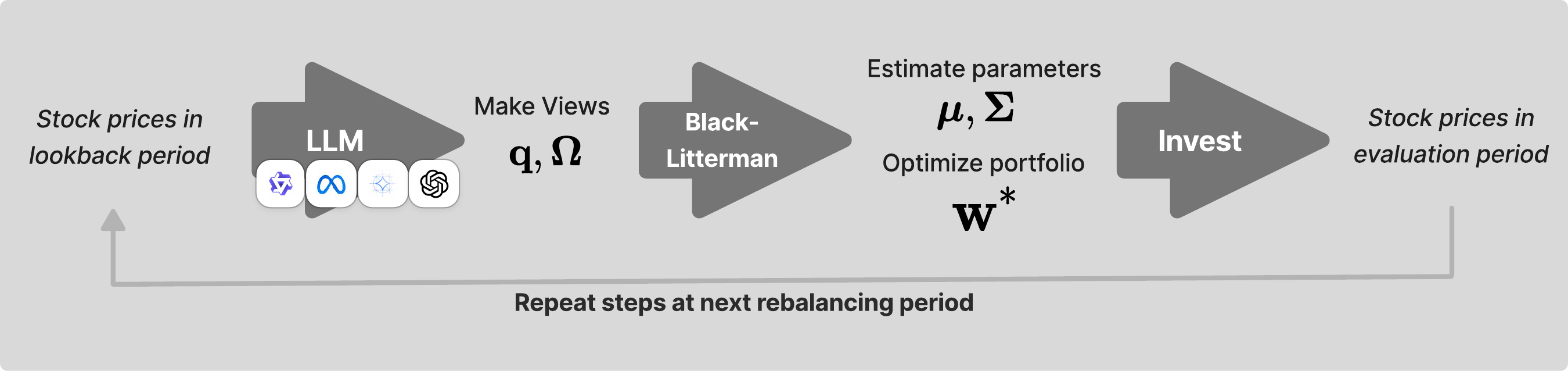}
    \caption{The Proposed Framework for Integrating LLM Views into Black-Litterman Model}
    \label{fig:model}
\end{figure*}

We first describe the financial dataset, then explain how LLMs are prompted to generate predictive views. Finally, we formalize the integration of these views and their underlying uncertainty into the Black-Litterman portfolio optimization framework.

\subsection{Data Description}

We use historical stock price data from the largest 50 S\&P 500 constituents by market capitalization as of March 26, 2025, sourced from Yahoo Finance. The total dataset spans 13 months, from June 2024 to June 2025. To ensure a robust evaluation and avoid overfitting, we divide the data into two distinct periods:

\begin{itemize}
    \item \textbf{Validation Period (3 months):} From June 2024 to August 2024. This period is used exclusively for tuning the hyperparameter $\tau$ of the Black-Litterman model.
    \item \textbf{Test Period (10 months):} From September 2024 to June 2025. This unseen data is used for the final performance evaluation of the portfolios.
\end{itemize}

The portfolio is rebalanced every two weeks. At each rebalancing date, the LLM is provided with the past two weeks of stock price data and company metadata to generate return predictions for the subsequent two weeks. This iterative process ensures that each portfolio construction is based on the most recent information available.

\subsection{Large Language Models}

We use LLMs to generate forward-looking views for the Black-Litterman model in an automated and scalable way. We detail this process, covering the selection of specific LLMs, our structured prompting technique for eliciting quantitative forecasts, and the method for measuring the uncertainty of these predictions.

\subsubsection{Selection of Large Language Models}

To incorporate predictive views, we employ four state-of-the-art models. Crucially, the knowledge cutoff for all selected models predates our entire evaluation period (Jun 2024–Jun 2025). This ensures that the models' internal knowledge does not contain information about future stock performance, thereby preventing any form of look-ahead bias. We use instruction-tuned models to ensure consistency in the generated return predictions, and use vLLM \cite{kwon2023efficient} for optimized inference.

\begin{table}[h]
    \centering
    \renewcommand{\arraystretch}{1.2}
    \begin{tabular}{llll}
        \hline
        \textbf{Model} & \textbf{Size} & \textbf{Release} & \textbf{Developer} \\
        \hline
        \texttt{Gemma-7B} \cite{team2024gemma}       & 7B     & Feb 2024 & Google DeepMind \\
        \texttt{Qwen-2-7B} \cite{yang2024qwen2technicalreport}      & 7B     & Jun 2024 & Alibaba Cloud \\
        \texttt{LLaMA-3.1-8B} \cite{dubey2024llama}   & 8B     & Jul 2024 & Meta \\
        \texttt{GPT-4o-mini} \cite{openai2024gpt4o}    & $\sim$8B \tnote{a} & Jul 2024 & OpenAI \\
        \hline
    \end{tabular}
    \begin{tablenotes}
        \footnotesize
        \item[a] Although the exact size of \texttt{GPT-4o-mini} has not been officially disclosed, \cite{abacha2024medec} estimate it to be within the 8 billion parameter range, making it comparable to the open-source models used in our study.
    \end{tablenotes}
    \caption{Summary of LLMs used in this study.}
    \label{tab:llm_summary}
\end{table}

\subsubsection{Structured Prompting for Predictive Views} \label{sec:prompt}


To elicit predictive views from the LLMs, we developed a structured prompting strategy that combines a detailed system prompt with a concise user prompt. This two-part structure is designed to guide the models toward generating quantitative, data-driven return forecasts while minimizing narrative or qualitative output. The full prompt structure is detailed in Appendix \ref{appendix:prompts}.

The system prompt sets the context, defines the task, and specifies the output format. It instructs the model to act as a financial analyst tasked with predicting the average daily return for a stock over the upcoming two-week period. It outlines the required analytical steps: analyzing the provided time-series data, considering the broader sector and market context, and incorporating company-specific information. Crucially, it constrains the output to a single floating-point number to ensure the response can be directly integrated into our quantitative framework.

The user prompt supplies the specific data for each stock at each rebalancing date. It includes four key pieces of information:
\begin{itemize}
    \item Stock Daily Returns: The stock's own daily returns from the past two weeks.
    \item Sector and Market Returns: The corresponding daily returns for the stock's GICS sector and the S\&P 500 index.
    \item Company Information: Metadata including the ticker, company name, GICS sector, and sub-industry.
\end{itemize}

\subsection{Integration with the Black-Litterman Model}

The overall workflow for integrating the predictive views and their uncertainty, generated by the LLMs, into the Black-Litterman framework is summarized in Figure~\ref{fig:model}. Our portfolio strategy adheres to a two-week rebalancing cycle.

At each rebalancing date, we first utilize the stock price data from the preceding two weeks (the lookback period). This data is fed into the LLM using the structured prompting technique detailed in Section \ref{sec:prompt} to generate predictive views for the subsequent two-week period. These resulting views, along with their corresponding confidence levels, are then integrated with the market equilibrium returns within the Black-Litterman model.

The model combines these inputs to estimate the posterior expected returns. Based on this updated estimation, portfolio optimization is performed to determine the new optimal asset allocation weights. This portfolio is then invested (held) according to these weights for the duration of the two-week evaluation period, at which point the cycle repeats. This entire process is executed iteratively at every rebalancing date throughout the test period.

\subsubsection{Overview of the Black-Litterman Model}

The Black-Litterman model is a portfolio optimization framework that combines market equilibrium returns derived from the Capital Asset Pricing Model (CAPM) \cite{sharpe1964capital} with investor views to generate more stable and intuitive asset allocations. Unlike traditional mean-variance optimization, which is highly sensitive to input parameters, the Black-Litterman model integrates external views by adjusting equilibrium returns in a Bayesian manner \cite{kim2024overview}.

In the standard Black-Litterman framework, the equilibrium return vector $\boldsymbol{\pi}$ is derived from market capitalization weights and serves as a prior estimate of expected returns for each stocks. Views on specific assets, denoted by a view vector $\textbf{q}$ and the corresponding confidence matrix $\boldsymbol{\Omega}$, are incorporated to adjust $\boldsymbol{\pi}$ and compute the posterior expected returns $\boldsymbol{\mu}$, which, along with the covariance matrix $\boldsymbol{\Sigma}$ of the stocks, are then used as input for portfolio optimization to get optimal weights $\mathbf{w}^*$ for stocks \cite{idzorek2007step}, as detailed in Appendix \ref{appendix:PO}.

That is, we solve for the posterior expected returns $\boldsymbol{\mu}$ from the Equation \ref{eq:main_equation}:

\begin{equation}
\mathbf{y} = \mathbf{X} \boldsymbol{\mu} + \boldsymbol{\varepsilon}, \quad \boldsymbol{\varepsilon} \sim \mathcal{N}(\mathbf{0}, \mathbf{V})
\label{eq:main_equation}
\end{equation}

\begin{equation}
\mathbf{y} = 
\begin{bmatrix}
\boldsymbol{\pi} \\
\mathbf{q}
\end{bmatrix}, \quad
\mathbf{X} = 
\begin{bmatrix}
\mathbf{I} \\
\mathbf{P}
\end{bmatrix}, \quad
\mathbf{V} = 
\begin{bmatrix}
\tau \mathbf{\Sigma} & \mathbf{0} \\
\mathbf{0} & \mathbf{\Omega}
\end{bmatrix}
\label{eq:matrix_definitions}
\end{equation}

where $\tau$ is a hyperparameter to balance confidence between the market equilibrium and the LLM-generated views. $\mathbf{\Sigma}$ is the covariance matrix of stock returns.

Our framework defines the remaining components using LLM outputs as described in Table \ref{tab:llm_views}.

\subsubsection{Incorporating LLM-Generated Views}

In this study, we extend the Black-Litterman model by leveraging LLM-generated predictive views instead of human-defined views. These views, extracted from multiple queries to LLMs, represent the expected average daily return for each asset and are incorporated into the model alongside their confidence levels.

When calculating the posterior expected returns $\boldsymbol{\mu}$, we can determine the equilibrium return vector $\boldsymbol{\pi}$ and the covariance matrix $\boldsymbol{\Sigma}$ of the stocks from historical return data. To integrate LLM-generated predictive views into the Black-Litterman model, we now define the view expectation vector $\textbf{q}$, the picking matrix $\mathbf{P}$, and the confidence matrix $\boldsymbol{\Omega}$ based on the LLM output. This process is described in Figure~\ref{fig:model}.
By structuring the LLM outputs in this manner, we systematically integrate them into the Black-Litterman framework.
The resulting posterior expected returns are then used in the portfolio optimization process.

\begin{table}[h!]
    \centering
    \begin{tabularx}{\columnwidth}{@{} l l X @{}} 
        \hline
        \textbf{Component} & \textbf{Symbol} & \textbf{Description} \\
        \hline
        View vector        & $\mathbf{q} \in \mathbb{R}^{k}$             & A vector containing the specific expected return forecasts for each of the \textit{k} views. \\
        Picking matrix     & $\mathbf{P} \in \mathbb{R}^{k \times n}$    & A matrix that maps the \textit{k} views to the corresponding \textit{n} assets in the portfolio. \\
        Confidence matrix  & $\boldsymbol{\Omega} \in \mathbb{R}^{k \times k}$ & A diagonal matrix quantifying the level of uncertainty for each of the \textit{k} views. \\
        \hline
    \end{tabularx}
    \caption{LLM-generated inputs for the Black-Litterman model}
    \label{tab:llm_views}
\end{table}

\paragraph{View Vector ($\mathbf{q} \in \mathbb{R}^{k}$)}
The vector $\mathbf{q}$ contains the expected return for each of the $k$ views. For each stock, the LLM is queried $N=100$ times, and the mean of these repeated responses constitutes the view for that stock. 

\paragraph{Picking matrix ($\mathbf{P} \in \mathbb{R}^{k \times n}$)}
In our framework, we formulate one absolute view for each individual stock, rather than relative views between assets.
The picking matrix specifies which of the \(n\) stocks correspond to each of the \(k\) views.
Since we generate an independent view for each of the $n$ stocks, the picking matrix $\mathbf{P}$ is an identity matrix ($k=n$), linking each view directly to a single asset.

\paragraph{Confidence Matrix ($\boldsymbol{\Omega} \in \mathbb{R}^{k \times k}$)}
This diagonal matrix captures the uncertainty of each view. The diagonal elements are populated with the variance of the $N$ repeated predictions for each stock. A higher variance signals greater uncertainty, automatically down-weighting that view's influence in the final blended return calculation.

\section{Experiment}


We conduct an empirical backtest to evaluate the performance of our proposed framework, which integrates LLM-generated views into the Black-Litterman model (BLM). To structure our analysis, we address the following key research questions:

\begin{itemize}
    \item[\textbf{RQ1}] \textit{(Performance)} How does the investment performance of portfolios constructed using our LLM-BLM framework compare to traditional benchmarks like EW and MVO in terms of absolute and risk-adjusted returns?
    \item[\textbf{RQ2}] \textit{(Style)} Do different LLMs exhibit distinct and characterizable `investment styles' in their predictive view distributions, and if so, how do they differ in terms of bias, dispersion, and conviction?
    \item[\textbf{RQ3}] \textit{(Attribution)} What is the relationship between an LLM's overarching predictive sentiment and its portfolio performance, and how does this relationship depend on the market regime?
\end{itemize}

The subsequent sections are structured to answer each of these questions in turn. Our evaluation follows a two-stage process: first, we optimize the model's key hyperparameter, $\tau$, which was introduced in Equation~\ref{eq:matrix_definitions}. Second, we assess the final portfolio performance on an unseen test period to address our research questions.

\subsection{Experimental Setup}

\begin{figure*}[t]
    \centering
    \includegraphics[width=0.9\textwidth]{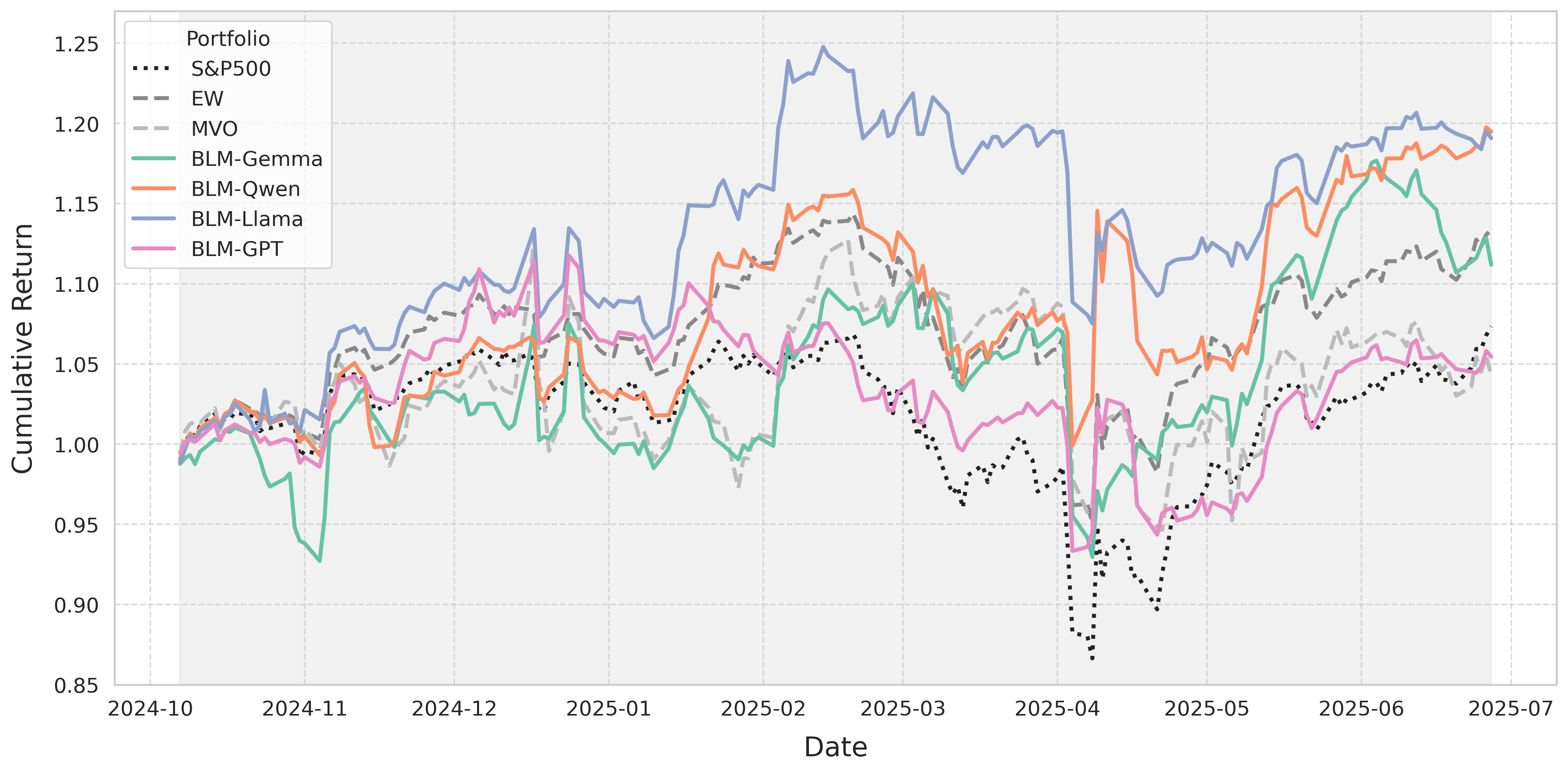}  
    \caption{
    Comparative Analysis of Cumulative Returns for BLM and Benchmark Portfolios. The results highlight the significant outperformance of the BLM-Qwen and BLM-Llama strategies, which consistently generated higher returns compared to both the market index and conventional quantitative models.}
    \label{fig:cumulative_returns}  
\end{figure*}

To ensure a comprehensive analysis, we evaluate four distinct LLM-enhanced Black--Litterman portfolios, each driven by views from a different model: \textbf{\texttt{BLM-Gemma}}, \textbf{\texttt{\textbf{\texttt{BLM-Qwen}}}}, \textbf{\texttt{\textbf{\texttt{BLM-Llama}}}}, and \textbf{\texttt{BLM-GPT}}. We compare their performance against three standard baselines:
\begin{itemize}
    \item \textbf{S\&P 500 Index (Market)}: A market-capitalization-weighted index that serves as the overall market benchmark.
    \item \textbf{Equally Weighted (EW)}: A simple yet robust baseline strategy that allocates equal capital to each asset, representing naive diversification.
    \item \textbf{Mean-Variance Optimization (MVO)}: The classic Markowitz portfolio model, which relies solely on historical return and covariance estimates and serves as a traditional quantitative benchmark.
\end{itemize}

The performance of each strategy is measured using a comprehensive suite of standard financial metrics. These include return indicators such as the Cumulative Annual Growth Rate (CAGR) alongside both daily (mean) and annualized (mean ann.) average returns; corresponding volatility (std); risk-adjusted returns (Sharpe); and key downside risk measures, including Maximum Drawdown (MDD), 95\% Value at Risk (VaR 95\%), and 95\% Conditional Value at Risk (CVaR 95\%).

Furthermore, to more realistically reflect the actual investment environment, transaction costs are incorporated into the backtest. Costs are calculated as 0.1\% of the portfolio weight changes at each rebalancing and are factored into the final performance. This setup allows us to rigorously test the practical value of integrating LLM-generated insights into a structured portfolio optimization framework.

\subsection{Hyperparameter Tuning}

The primary hyperparameter in our framework is $\tau$, which, as introduced in the variance-covariance matrix $\mathbf{V}$ in Equation~\eqref{eq:matrix_definitions}, \textit{balances the confidence between the market equilibrium prior and the LLM-generated views}. A higher value of $\tau$ indicates less confidence in the prior (market equilibrium returns), thus placing more weight on the LLM's views. Instead of relying on a fixed, arbitrary value, we tune $\tau$ on the validation dataset. This is a crucial step to find the optimal balance between market priors and the specific characteristics of each LLM's views for our asset universe. We employ a two-step process for this tuning.

\subsubsection{Step 1: Heuristic Estimation of Initial \texorpdfstring{$\tau$}{tau}}

First, we derive a heuristic initial estimate for $\tau$, denoted as $\tau_{\text{init}}$. This value is designed to dynamically ground our model's confidence based on the relative uncertainty between the LLM's views and the market's historical volatility. We calculate $\tau_{\text{init}}$ by averaging the ratio of these uncertainties across all rebalancing periods ($t$) in the validation set ($T_{val}$):
\begin{equation}
\tau_{\text{init}} = \frac{1}{|T_{val}|} \sum_{t \in T_{val}} \frac{\text{mean}(\boldsymbol{\Omega}_t)}{\text{mean}(\boldsymbol{\Sigma}_t)}
\label{eq:tau_init}
\end{equation}
where $\boldsymbol{\Omega}_t$ is the confidence matrix of the LLM views and $\boldsymbol{\Sigma}_t$ is the covariance matrix of asset returns at rebalancing period $t$. The $\text{mean}(\cdot)$ function computes the mean of all elements in the given matrix. This approach ensures our initial hyperparameter is directly informed by the relative uncertainties observed during the validation period.

\subsubsection{Step 2: Grid Search for Optimal \texorpdfstring{$\tau$}{tau}}

Using $\tau_{\text{init}}$ as a center point, we perform a grid search over a range of values, $\mathcal{T}$, defined as:
$$
\mathcal{T} = \{0.5\tau_{\text{init}}, 0.75\tau_{\text{init}}, \tau_{\text{init}}, 1.25\tau_{\text{init}}, 1.5\tau_{\text{init}}\}
$$
For each $\tau \in \mathcal{T}$, we run a full backtest on the validation period and calculate the resulting Sharpe ratio. The $\tau$ value that yields the highest Sharpe ratio is selected as the optimal hyperparameter, $\tau^*$.

The optimal hyperparameter, $\tau^*$, found during the validation phase is then used for the final evaluation. We run a single backtest over the test period (Sep 2024-Jun 2025), rebalancing the portfolio every two weeks using this fixed $\tau^*$.

\subsection{RQ1: Portfolio Performance}

\begin{table}[h]

    \centering
    \renewcommand{\arraystretch}{1.2}
    \sisetup{
        round-mode=places,
        round-precision=4,
        detect-weight=true,
        table-format=-1.4
    }
    \footnotesize 
    \setlength{\tabcolsep}{2.8pt} 
    \begin{tabularx}{\columnwidth}{ 
        >{\RaggedRight}X  
        S[table-format=1.4]
        S[table-format=1.4]
        S[table-format=-1.4]
        S[table-format=1.4]
        S[table-format=1.4]
        S[table-format=-1.4]
    }
        \hline
        {Metric} & {EW} & {MVO} & {\texttt{BLM-Gemma}} & {\texttt{BLM-Qwen}} & {\texttt{BLM-Llama}} & {\texttt{BLM-GPT}} \\
        \hline
        CAGR \(\uparrow\) & 0.1907 & 0.0607 & 0.1590 & \bfseries 0.2811 & \ul{0.2751} & 0.0768 \\
        mean \(\uparrow\) & 0.0008 & 0.0004 & 0.0007 & \bfseries 0.0011 & \ul{0.0010} & 0.0004 \\
        std \(\downarrow\) & \bfseries 0.0122 & 0.0179 & 0.0156 & 0.0152 & \ul{0.0124} & 0.0132 \\
        Sharpe \(\uparrow\) & 0.0563 & 0.0176 & 0.0402 & \ul{0.0669} & \bfseries 0.0774 & 0.0228 \\
        mean (ann.) \(\uparrow\) & 0.1932 & 0.0994 & 0.1784 & \bfseries 0.2763 & \ul{0.2627} & 0.0958 \\
        std (ann.) \(\downarrow\) & \bfseries 0.1938 & 0.2841 & 0.2481 & 0.2413 & \ul{0.1975} & 0.2093 \\
        Sharpe (ann.) \(\uparrow\) & 0.8937 & 0.2793 & 0.6386 & \ul{1.0624} & \bfseries 1.2286 & 0.3619 \\
        MDD \(\uparrow\) & -0.1688 & -0.1620 & -0.1551 & \bfseries -0.1375 & \ul{-0.1383} & -0.1649 \\
        VaR95\% \(\uparrow\) & -0.0171 & -0.0259 & \ul{-0.0168} & -0.0182 & -0.0169 & \bfseries -0.0157 \\
        CVaR95\% \(\uparrow\) & \bfseries -0.0280 & -0.0448 & -0.0372 & -0.0320 & \ul{-0.0283} & -0.0312 \\
        \hline
    \end{tabularx}
    \caption{
    Out-of-sample Investment Performance of Portfolios
    }
    \label{tab:vertical_performance_comparison}
\end{table}

\begin{figure*}[b]
    \centering
    \includegraphics[width=1.0\textwidth]{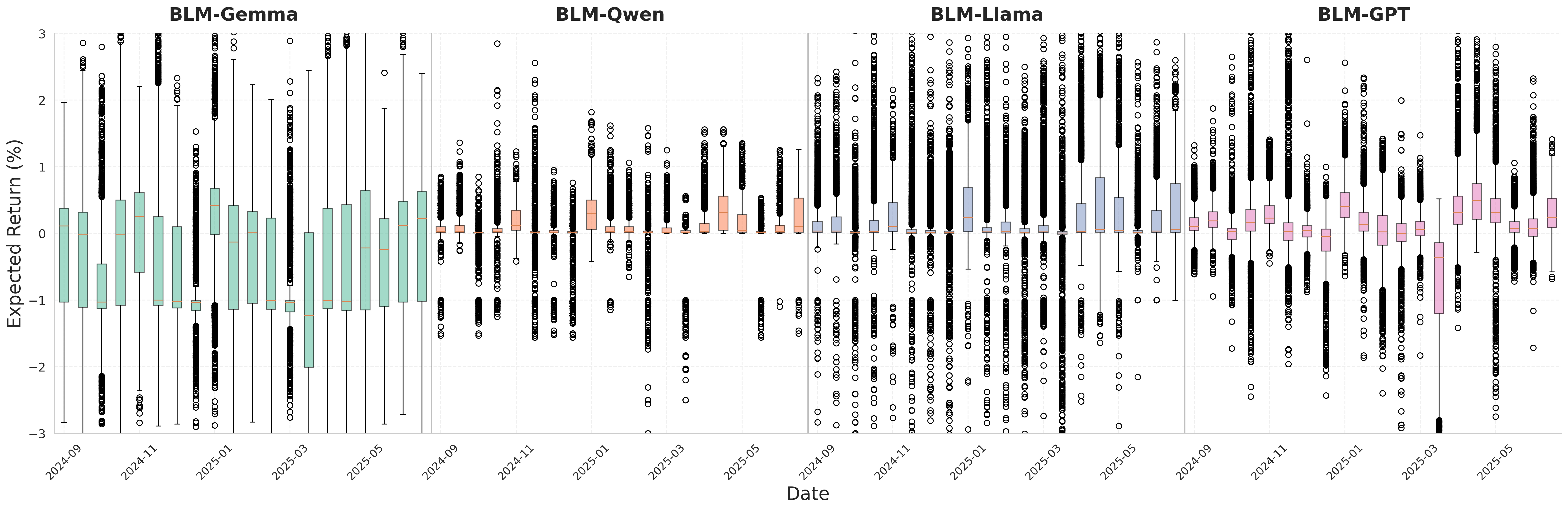} 
    \caption{
    LLM-generated Views Over Time at Rebalancing Intervals. The figure visualizes the distribution of all generated return forecasts for each LLM at every two-week rebalancing interval. 
    }
    \label{fig:boxplot} 
\end{figure*}

Table~\ref{tab:vertical_performance_comparison} and Figure~\ref{fig:cumulative_returns} summarize the comparative performance analysis of four proposed BLM-based models (\textbf{\texttt{BLM-Gemma}}, \textbf{\texttt{BLM-Qwen}}, \textbf{\texttt{BLM-Llama}}, and \textbf{\texttt{BLM-GPT}}) against two traditional benchmarks: the EW and MVO portfolios. The evaluation encompasses various metrics, including absolute returns, risk-adjusted returns, and downside risk.

The analysis reveals a strong performance from the BLM-based models, particularly \textbf{\texttt{BLM-Qwen}} and \textbf{\texttt{BLM-Llama}}. \textbf{\texttt{BLM-Qwen}} achieved the highest CAGR of 0.2811, closely followed by \textbf{\texttt{BLM-Llama}} at 0.2751. Both significantly outperformed the traditional benchmarks, EW (0.1907) and MVO (0.0607).

In terms of risk-adjusted returns, measured by the annualized Sharpe Ratio, Sharpe (ann.), \textbf{\texttt{BLM-Llama}} delivered the top performance with a ratio of 1.2286. \textbf{\texttt{BLM-Qwen}} secured the second-highest position at 1.0624. This indicates that both models generated superior excess returns per unit of risk, far exceeding the benchmarks EW (0.8937) and MVO (0.2793).

Regarding volatility, std (ann.), the EW portfolio was the most stable with a value of 0.1938, underscoring the benefits of simple diversification. However, \textbf{\texttt{BLM-Llama}} was a very close second with a volatility of 0.1975, indicating effective risk management. In downside risk, the BLM models also excelled. \textbf{\texttt{BLM-Qwen}} posted the lowest MDD at -0.1375, with \textbf{\texttt{BLM-Llama}} following closely at -0.1383, both demonstrating significantly better capital preservation than the benchmarks during downturns.

The other LLM-based strategies showed mixed results. The \textbf{\texttt{BLM-Gemma}} model delivered a modest positive CAGR of 0.1590, while the \textbf{\texttt{BLM-GPT}} model lagged with a CAGR of 0.0768, performing similarly to the MVO benchmark in terms of returns.

In conclusion, the framework successfully translated LLM-generated views into high-performing portfolios. The \textbf{\texttt{BLM-Qwen}} and \textbf{\texttt{BLM-Llama}} models, in particular, demonstrated exceptional and distinct strengths: \textbf{\texttt{BLM-Qwen}} excelled in generating the highest absolute returns (CAGR), while \textbf{\texttt{BLM-Llama}} delivered the best risk-adjusted performance (Sharpe) and showcased strong risk control. Both models decisively outperformed traditional benchmarks and other LLM strategies, highlighting the framework's effectiveness.

\subsection{RQ2: Investment Styles}

To further investigate how LLMs formulated asset views, we analyzed the distribution of their individual return predictions. In our setup, each LLM repeatedly generated predictions per stock at each rebalancing point. Figure~\ref{fig:boxplot} displays a time-series of these distributions. Specifically, each individual boxplot on the chart represents the entire population of views ($N$ predictions × $n$ assets) generated by a model for that single rebalancing period. This visualization method reveals how the overall distribution of views—encompassing both cross-sectional differences across assets and the inherent uncertainty within the predictions—evolves over time. The y-axis is clipped to a range of -3\% to 3\% for clarity, and the summary statistics in Table~\ref{tab:view_statistics} provide details on the aggregate distribution across all periods.

The most striking characteristic is the distribution of views from \textbf{\texttt{BLM-Llama}}. The boxplots and summary statistics reveal an exceptionally wide dispersion of predictions (Std: 0.6903) with extreme outliers (Min: -57.99\%, Max: 91.70\%). This indicates that \textbf{\texttt{BLM-Llama}} generated highly differentiated and high-conviction views, confidently identifying both strong positive and strong negative return opportunities. Despite the extreme negative outliers, the overall distribution maintained the highest positive mean (0.1786), suggesting its positive views were particularly impactful. This "sharp" differentiation between assets appears to be a key driver of its strong portfolio performance.

In stark contrast, \textbf{\texttt{BLM-Gemma}} and \textbf{\texttt{BLM-Qwen}} represent two opposite poles of view generation. \textbf{\texttt{BLM-Gemma}} consistently exhibited a strong pessimistic bias; its median view was almost always negative, a fact confirmed by its negative overall mean of -0.3847. This persistent negative outlook on the market likely led to suboptimal asset selection. Conversely, \textbf{\texttt{BLM-Qwen}} adopted a cautiously optimistic stance. Its views were the most tightly clustered around a stable, slightly positive median, reflected in the lowest standard deviation (0.2872) among all models. This suggests a consistent, low-conviction strategy that avoids extreme bets.

The \textbf{\texttt{BLM-GPT}} model can be characterized as a middle ground. Its view distribution was more dispersed than \textbf{\texttt{BLM-Qwen}}'s but less biased than \textbf{\texttt{BLM-Gemma}}'s, with a moderately positive mean (0.1347) and standard deviation (0.4805). Its predictions were less stable than Qwen's but avoided the persistent pessimism of Gemma.

In conclusion, these results clearly show that the "style" of view generation is critical. The final portfolio outcomes are a direct reflection of the distinct characteristics of each model's predictive distribution. \textbf{\texttt{BLM-Llama}}'s high-conviction differentiation, \textbf{\texttt{BLM-Qwen}}'s stable and cautious optimism, and \textbf{\texttt{BLM-Gemma}}'s pervasive pessimism are not just statistical artifacts but are the fundamental drivers that explain their divergent investment performance.

\subsection{RQ3: Performance Attribution to Predictive Sentiment}

\begin{figure*}[h]
    \centering
    \includegraphics[width=0.9\textwidth]{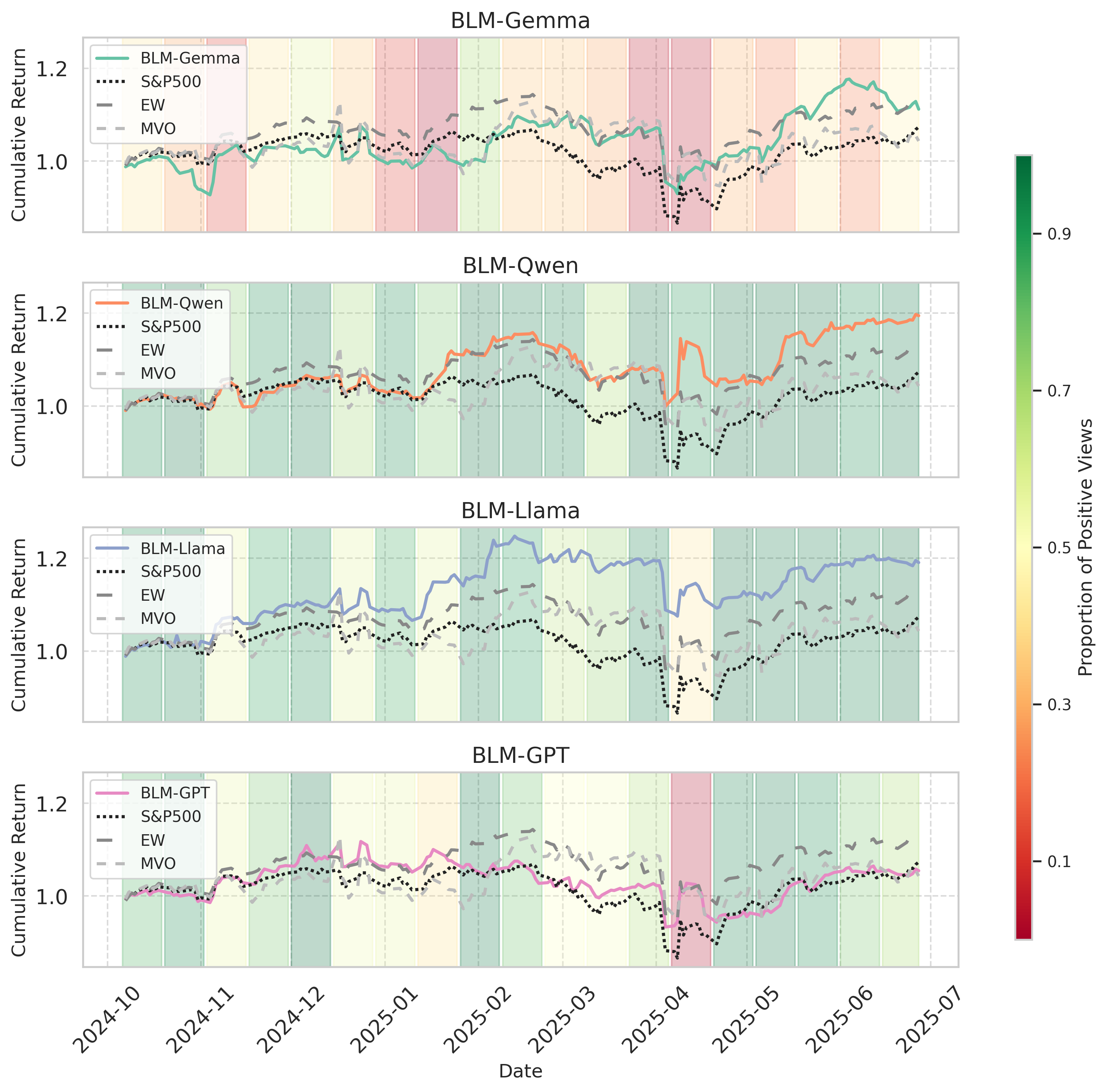} 
    \caption{Correlation Between LLM Predictive Sentiment and Portfolio Performance. This figure compares the cumulative return of each LLM-driven portfolio (line graphs) with the underlying sentiment of the LLM's views at each rebalancing point (background color). 
    }
    \label{fig:proportion} 
\end{figure*}

Figure~\ref{fig:proportion} visualizes the cumulative returns of each portfolio model, with the background color illustrating the proportion of positive views among all generated predictions at each rebalancing point. The color scale ranges from pessimistic (red, a low proportion of positive views) to optimistic (green, a high proportion). This analysis reveals a direct link between a model's overarching market sentiment and its investment performance, contingent on the market conditions of the test period.

\begin{enumerate}

    \item \textbf{Alignment of Optimism with Market Trend in Top Performers:} The top-performing models, \textbf{\texttt{BLM-Llama}} and \textbf{\texttt{BLM-Qwen}}, both demonstrate a clear synchronization between a consistently optimistic market outlook and strong portfolio returns. Their backgrounds remain predominantly green, indicating a high proportion of positive views. During the observed test period, which featured a generally upward market trend (as indicated by the S\&P500 benchmark), this stable, bullish conviction proved to be the correct strategic stance. It allowed the models to effectively participate in market updrafts and generate significant alpha.
    
    \item \textbf{Misaligned Sentiment in Underperformers:} In stark contrast, \textbf{\texttt{BLM-Gemma}}'s underperformance is clearly linked to a persistent pessimistic bias. Its background is dominated by red and yellow shades, signifying a low proportion of positive views. This consistently bearish outlook was fundamentally misaligned with the market's overall upward trajectory during the period, causing the model to miss key appreciation opportunities and leading to a decline in cumulative return. Separately, \textbf{\texttt{BLM-GPT}}'s mediocre performance appears driven by an unstable market view. Its frequent vacillation between bullish and bearish sentiment suggests an inability to commit to the prevailing market trend, resulting in performance that failed to meaningfully outperform the benchmarks.

    \item \textbf{Alignment with Market Regime as the Key Differentiator:} In summary, this visual analysis confirms that a model's performance is critically linked to how well its overarching sentiment aligns with the market regime of the test period. The success of \textbf{\texttt{BLM-Llama}} and \textbf{\texttt{BLM-Qwen}} was rooted in their stable, optimistic view-generation process, which was the appropriate stance for the generally bullish environment observed. Conversely, \textbf{\texttt{BLM-Gemma}}'s failure can be attributed to its pervasive pessimism being out of sync with the market, while \textbf{\texttt{BLM-GPT}}'s middling results were a product of its inconsistent outlook. This highlights that for this framework to succeed, the underlying LLM must generate a consistent and, crucially, \textit{correctly calibrated} conviction about the overall market direction for a given period.

\end{enumerate}

\section{Conclusion}

This study proposed and validated a systematic framework for translating LLM predictions and their inherent uncertainty into actionable views for the Black-Litterman model. Our empirical backtest demonstrated that this approach generates portfolios, particularly from \texttt{BLM-Llama} and \texttt{BLM-Qwen}, that significantly outperform traditional benchmarks. Crucially, our analysis revealed that performance is a direct consequence of each LLM's distinct view generation `style'—such as \texttt{BLM-Llama}'s high-conviction differentiation versus \texttt{BLM-Qwen}'s stable optimism. The key implication is that the choice of an LLM is akin to selecting an investment manager with a specific bias, whose success is contingent on its alignment with the prevailing market regime. Future research could extend this work by incorporating financial text into the view generation process or by developing methods to dynamically select models based on expected market conditions.

\begin{acks}
This research was supported by \textbf{Brian Impact Foundation}, a non-profit organization dedicated to the advancement of science and technology for all.
\end{acks}

\bibliographystyle{ACM-Reference-Format}
\bibliography{_main}

\clearpage
\appendix

\section{Prompt Design Details} \label{appendix:prompts}

To clarify the temporal context for the return prediction task, we include an explicit reference date in the system prompt. This date represents the point in time at which the LLM is expected to provide its analysis—positioned between the past time series data and the future return period. By anchoring the prediction task to a specific date, we help the model better interpret the directionality of the input data and generate temporally coherent forecasts.

To ensure consistent and parseable responses from the LLMs, we adopt a structured output format using the JSON schema supported by the chat completions API endpoint. This approach standardizes the outputs across all LLMs and ensures that each model provides consistently structured responses across repeated queries, reducing the likelihood of formatting errors and facilitating reliable downstream processing.

To enhance the models' numerical sensitivity, all daily return inputs are scaled by 100 (i.e., presented in percentage terms). For instance, a daily return of -0.0036 is provided as -0.36. This scaling prevents issues arising from the small magnitude of raw daily returns, which can be difficult for LLMs to differentiate and process effectively. This approach is analogous to using basis points in finance to handle small numerical values. By providing a rich, multi-faceted data context and clear instructions, the prompts enable the LLMs to generate informed, quantitative views for portfolio construction.

\begin{figure}[h]
\centering
    \begin{tcolorbox}[colframe=blue!80!black, colback=blue!10!white, coltitle=white, title=System Prompt, width=1\columnwidth]
    \small
    You are providing analysis on \texttt{{\{\{DATE\}\}}}. Predict the average daily return for the next two weeks based on the information provided about a stock's past performance.
    
    You will receive the following inputs:
    \begin{itemize}
        \item \textbf{Daily Returns}: The stock's daily returns, a time-series from the past two weeks.
        \item \textbf{Company Sector}: The company's GICS sector classification.
        \item \textbf{Sector Returns}: The company sector's daily returns, a time-series from the past two weeks.
        \item \textbf{Market Returns}: The S\&P 500's daily returns, a time-series from the past two weeks.
        \item \textbf{Company Information}:
            \begin{itemize}
                \item Ticker: The stock symbol.
                \item Company Name: The name of the company.
                \item GICS Sector: The Global Industry Classification Standard sector.
                \item GICS Sub-Industry: The sub-industry classification.
            \end{itemize}
    \end{itemize}
    
    \# Steps
    
    \begin{enumerate}
        \item \textbf{Analyze the Time-Series Data}: Review the historical daily returns to identify patterns, trends, or anomalies that may affect future performance.
        \item \textbf{Consider Sector Performance}: Analyze how the market and the sector's performance might influence the stock's future returns.
        \item \textbf{Incorporate Company Information}: Use the details from the GICS sector and sub-industry, along with the company symbol and name, to contextualize the predicted performance within its industry.
        \item \textbf{Predict Future Returns}: Estimate the average daily returns for the next two weeks based on the analysis of available data.
    \end{enumerate}
    
    \# Output Format
    
    \begin{quote}
    Return a single float value that represents the predicted average daily return for the stock over the next two weeks, without any additional commentary or explanation.
    \end{quote}
    
    \# Notes
    
    \begin{itemize}
        \item Ensure the prediction considers the quantified data from the time-series.
        \item Make calculations based on statistical trends from daily returns data.
        \item Pay attention to the trends within both the stock's daily returns and the market's return data.
        \item Consider the relevance of the company's sector to refine your predictions.
        \item Make calculations without additional interpretation or commentary.
    \end{itemize}
    
    \end{tcolorbox}
    \caption{The structure of the system prompt used to elicit return forecasts from the LLMs.}
    \label{fig:system_prompt}
\end{figure}


\begin{figure}[h]
\centering
    \begin{tcolorbox}[colframe=green!60!black, colback=green!5!white, coltitle=white, title=User Prompt, width=1\columnwidth]
    \small
    \ttfamily 
    \textbf{Daily Returns:} [-1.17, -0.92, -2.31, -0.36, -3.02, 2.53, 0.1, -0.23, 0.45]
    
    \vspace{1mm}
    \textbf{Company Sector:} Information Technology
    
    \vspace{1mm}
    \textbf{Sector Returns:} [-0.14, -0.38, -1.88, 0.03, -1.1, 2.38, -0.41, 0.17, -0.31]
    
    \vspace{1mm}
    \textbf{Market Returns:} [0.09, -0.39, -1.61, -0.04, -0.67, 2.05, -0.56, 0.4, -0.01]
    
    \vspace{1mm}
    \textbf{Company Information:}
    
    \hspace{5mm} \textbf{Ticker:} AAPL
    
    \hspace{5mm} \textbf{Company Name:} Apple Inc.
    
    \hspace{5mm} \textbf{GICS sector:} Information Technology
    
    \hspace{5mm} \textbf{GICS sub-industry:} Technology Hardware, Storage \& Peripherals
    
    \end{tcolorbox}
    \caption{An example user prompt providing data for Apple Inc. (AAPL) on a specific rebalancing date.}
    \label{fig:user_prompt}
\end{figure} 

\section{Portfolio Optimization} \label{appendix:PO}

In this study, we implement two distinct portfolio optimization approaches: one based on the views generated by LLMs and the other using traditional mean-variance optimization as a baseline. The objective is to compare the performance of LLM-based portfolios against a benchmark portfolio optimized using conventional techniques.

\subsection{LLM-Enhanced Black-Litterman Portfolio}  
The first approach utilizes the posterior returns derived from the views generated by the LLMs and covariance matrix $\boldsymbol{\Sigma}$. The portfolio optimization problem is formulated as a mean-variance optimization problem, where the goal is to maximize the expected return of the portfolio while minimizing its risk.

The optimization problem is expressed as:

\[
\min_{\mathbf{w}} \left( \mathbf{w}^T \Sigma \mathbf{w} - \lambda \cdot \mathbf{w}^T \mathbf{r} \right)
\]

where:
\begin{itemize}
    \item \(\mathbf{w}\) is the vector of portfolio weights,
    \item \(\mathbf{r}\) is the vector of expected returns (posterior returns) from the LLM views,
    \item \(\Sigma\) is the covariance matrix of returns,
    \item \(\lambda\) is the risk aversion parameter, which determines the trade-off between return and risk.
\end{itemize}

In this formulation, \(\mathbf{w}^T \mathbf{r}\) represents the portfolio return, and \(\mathbf{w}^T \Sigma \mathbf{w}\) represents the portfolio risk (variance). The objective is to maximize portfolio return while minimizing the risk, adjusted by the risk aversion parameter \(\lambda\), which we set as 0.1. The optimization is subject to the constraints:
\begin{itemize}
    \item The sum of the portfolio weights equals 1 ($\sum w_i = 1$),
    \item No shorting is allowed, meaning each weight is non-negative ($w_i \geq 0$ for all $i$).
\end{itemize}
The optimal portfolio weights are obtained by minimizing the objective function.

\subsection{Baseline Portfolio}  
For comparison, MVO is optimized using the traditional mean-variance framework, which is based on historical returns and the covariance matrix of asset returns. The mean-variance optimization problem aims to maximize the expected return while minimizing the portfolio’s risk.

The optimization problem is formulated as:

\[
\min_{\mathbf{w}} \left( \mathbf{w}^T \Sigma \mathbf{w} - \lambda \cdot \mathbf{w}^T \mathbf{r} \right)
\]

where:
\begin{itemize}
    \item \(\mathbf{w}^T \Sigma \mathbf{w}\) represents the portfolio risk (variance),
    \item \(\mathbf{w}^T \mathbf{r}\) represents the portfolio return,
\end{itemize}

The portfolio risk (variance) and portfolio return are defined similarly in both optimization problems, with \(\mathbf{w}^T \Sigma \mathbf{w}\) representing risk and \(\mathbf{w}^T \mathbf{r}\) representing return. The optimization problem is subject to the same constraints:
\begin{itemize}
    \item The sum of the portfolio weights equals 1 ($\sum w_i = 1$),
    \item No shorting is allowed, meaning each weight is non-negative ($w_i \geq 0$ for all $i$).
\end{itemize}
The result of this optimization provides the portfolio weights for the baseline portfolio.

\subsection{Detailed Model-Level Statistics}

\subsubsection{View Statistics}

Table~\ref{tab:view_statistics} provides the summary statistics for the entire population of raw views generated by each LLM throughout the test period. These statistics reveal the distinct predictive "styles" of the models that were discussed in the main analysis. For instance, \textbf{\texttt{BLM-Llama}} is characterized by the highest mean return but also the largest standard deviation and the widest min-max range, indicating a high-conviction, high-dispersion approach. In contrast, \textbf{\texttt{BLM-Qwen}} shows the lowest standard deviation, reflecting a more cautious and stable view generation process, while \textbf{\texttt{BLM-Gemma}} is notable for its negative mean, confirming a persistent pessimistic bias. These underlying statistical properties are the root cause of the performance differences analyzed in the main text.

\begin{table}[h!]
    \centering
    \caption{Summary Statistics of LLM-Generated Views}
    \label{tab:view_statistics}
    \renewcommand{\arraystretch}{1.2}
    \sisetup{
        round-mode=places,
        round-precision=4,
        table-format=-2.4 
    }
    \small
    \setlength{\tabcolsep}{5pt}
    \begin{tabular}{l S S S S}
    \toprule
    \textbf{Metric} & {\textbf{\texttt{BLM-Gemma}}} & {\textbf{\texttt{BLM-Qwen}}} & {\textbf{\texttt{BLM-Llama}}} & {\textbf{\texttt{BLM-GPT}}} \\
    \midrule
    \textbf{Count}      & \multicolumn{1}{r}{100000} & \multicolumn{1}{r}{100000} & \multicolumn{1}{r}{100000} & \multicolumn{1}{r}{100000} \\
    \textbf{Mean}       & -0.3847       & 0.1007        & 0.1786        & 0.1347        \\
    \textbf{Std}        & 0.9711        & 0.2872        & 0.6903        & 0.4805        \\
    \textbf{Min}        & -22.3000      & -4.7500       & -57.9913      & -6.9250       \\
    \textbf{25\%}       & -1.1200       & 0.0100        & 0.0075        & -0.0244       \\
    \textbf{Median}     & -0.6600       & 0.0300        & 0.0270        & 0.1000        \\
    \textbf{75\%}       & 0.3200        & 0.1200        & 0.2000        & 0.3190        \\
    \textbf{Max}        & 4.8100        & 3.1400        & 91.6982       & 6.1140        \\
    \bottomrule
    \end{tabular}
\end{table} 

\subsubsection{Analysis of Forecasting Errors} %

Table~\ref{tab:prediction_errors} presents an analysis of the forecasting accuracy for the views generated by each LLM and for the traditional estimate used by the MVO benchmark. The table quantifies the prediction error by comparing the expected returns—either generated by an LLM or, in the case of MVO, estimated from the historical average of past returns—against the actual realized returns of the subsequent period. This error is measured using three standard metrics where lower values are better: Mean Squared Error (MSE), Root Mean Squared Error (RMSE), and Mean Absolute Error (MAE).

The results reveal a strong correspondence between a model's forecasting accuracy and its ultimate portfolio performance, as detailed in Table~\ref{tab:vertical_performance_comparison}. \textbf{\texttt{BLM-Qwen}} and \textbf{\texttt{BLM-Llama}}, which demonstrated the highest predictive accuracy by achieving the lowest errors across all three metrics, also generated the portfolios with the best absolute and risk-adjusted returns (CAGR and Sharpe Ratio respectively). This provides direct evidence that their outperformance was rooted in a superior underlying ability to forecast market returns. Conversely, the MVO benchmark, relying on naive historical averages, exhibited high prediction errors and correspondingly poor investment results. Interestingly, while \textbf{\texttt{BLM-Gemma}} also showed high forecasting errors—even worse than MVO's—its final portfolio outperformed both MVO and \textbf{\texttt{BLM-GPT}} in terms of CAGR. This suggests that within the Black-Litterman framework, the nature of the views can be as important as their raw accuracy; the consistent (though pessimistic) bias of Gemma may have provided a more stable, albeit flawed, signal for the optimizer to work with.

\begin{table}[h!]
    \centering
    \caption{Comparison of Prediction Errors}
    \label{tab:prediction_errors}
    \renewcommand{\arraystretch}{1.2}
    \sisetup{
        round-mode=places,
        round-precision=4,
        detect-weight=true,
        table-format=1.4 
    }
    \small
    \setlength{\tabcolsep}{5pt}
    \begin{tabular}{l S S S S S}
    \toprule
    \textbf{Metric} & {\textbf{MVO}} & {\textbf{\texttt{BLM-Gemma}}} & {\textbf{\texttt{BLM-Qwen}}} & {\textbf{\texttt{BLM-Llama}}} & {\textbf{\texttt{BLM-GPT}}} \\
    \midrule
    MSE $\downarrow$  & 0.9376 & 1.2373 & \bfseries 0.5125 & \ul{0.5288} & 0.6505 \\
    RMSE $\downarrow$ & 0.9683 & 1.1123 & \bfseries 0.7159 & \ul{0.7272} & 0.8066 \\
    MAE $\downarrow$  & 0.6989 & 0.8608 & \bfseries 0.5168 & \ul{0.5281} & 0.5702 \\
    \bottomrule
    \end{tabular}
\end{table} 

\subsection{Analysis of Portfolio Allocation Dynamics}

\begin{figure*}[ht]
    \centering
    \includegraphics[width=1\linewidth, height=1\textheight, keepaspectratio]{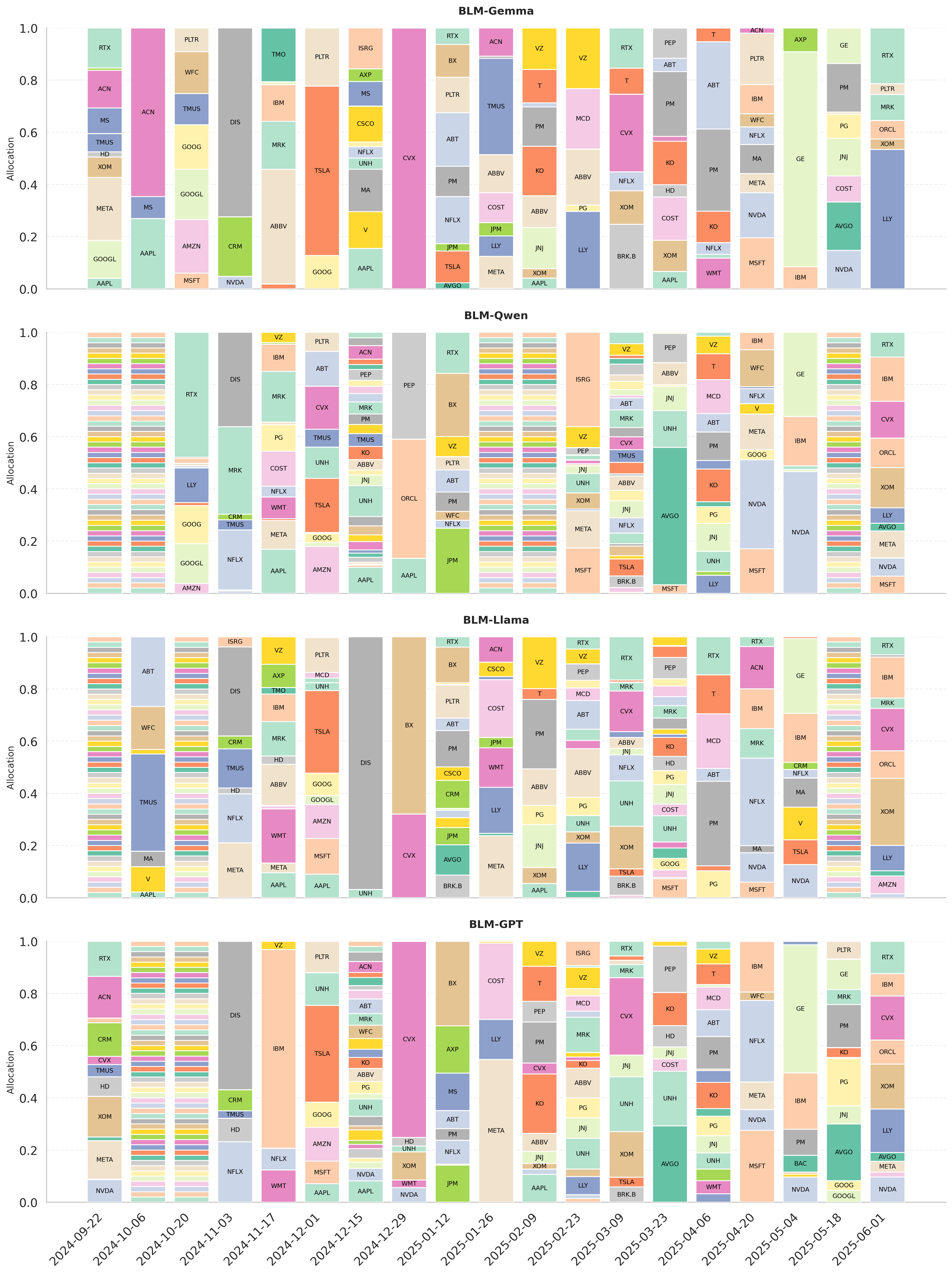}
    \caption{Time-series of Asset Allocation Weights for \textbf{BLM Portfolios}}
    \label{fig:weights_change_LLM}
\end{figure*}

\begin{figure*}[ht]
    \centering
    \includegraphics[width=1\linewidth, height=1\textheight, keepaspectratio]{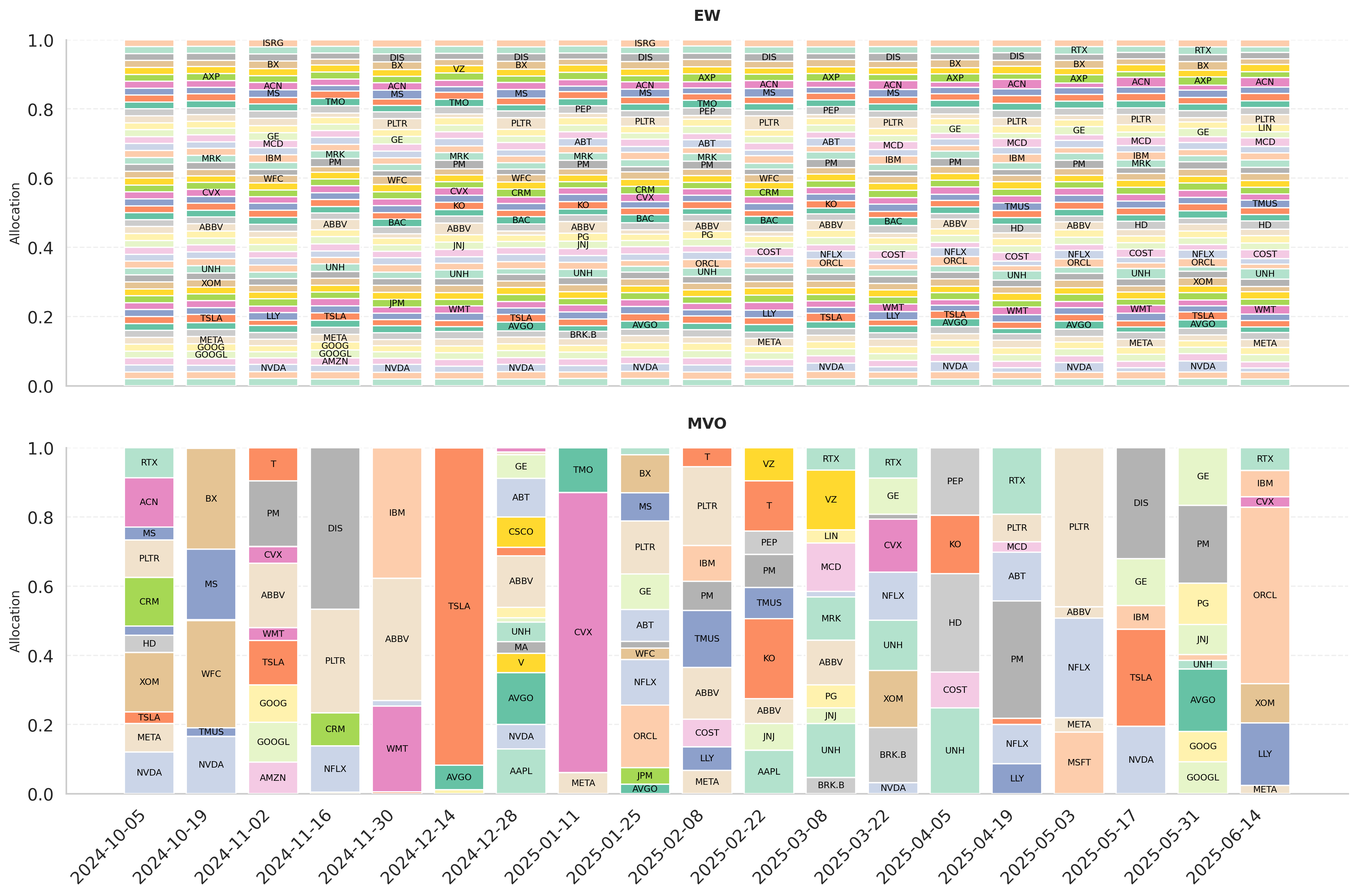}
    \caption{Time-series of Asset Allocation Weights for \textbf{Baseline Portfolios}}
    \label{fig:weights_change_baselines}
\end{figure*}

Figures~\ref{fig:weights_change_LLM} and~\ref{fig:weights_change_baselines} illustrate the temporal evolution of portfolio allocations for the LLM-driven and baseline strategies, respectively. 

Figure~\ref{fig:weights_change_LLM} details the asset weights at each rebalancing date for the portfolios constructed using our proposed framework. The allocations shown are dynamically adjusted based on the predictive views generated by each respective LLM. For comparison, Figure~\ref{fig:weights_change_baselines} displays the allocations for the baseline portfolios. These include the static, uniformly distributed EW strategy and the MVO portfolio, whose weights are derived solely from historical data. 

\end{document}